\documentclass[twocolumn,aps,prl,superscriptaddress]{revtex4-2}
\usepackage[utf8]{inputenc}
\usepackage{epsf,graphicx}
\usepackage{multirow}
\usepackage{amsmath,gensymb,amssymb}
\usepackage{lipsum}
\usepackage{dcolumn}
\usepackage{bm}
\usepackage{hyperref}
\usepackage{natbib}
\usepackage{color}
\usepackage{color}
\usepackage{ulem}

\begin{document}

	\title{Plane-Symmetric Capillary Turbulence: Five-Wave Interactions}

	\author{E.A. Kochurin}
	\email{kochurin@iep.uran.ru}
	\affiliation{Institute of Electrophysics, Ural Division, Russian Academy of Sciences, Yekaterinburg, 620016 Russia}
	\affiliation{Skolkovo Institute of Science and Technology, 121205, Moscow, Russia}
	
	\author{P.A. Russkikh}
	\affiliation{Institute of Electrophysics, Ural Division, Russian Academy of Sciences, Yekaterinburg, 620016 Russia}
	
	\begin{abstract}
The theory of isotropic capillary turbulence was developed in the late 1960s by Zakharov and Filonenko. 
To date, the analytical solution of the kinetic equation describing the stationary transfer of energy to small scales due to three-wave resonant interactions, called the Zakharov-Filonenko spectrum, has been confirmed with high accuracy.
However, in the case of strong anisotropy in wave propagation, where all waves are collinear, the situation changes significantly. In such a degenerate geometry, the conditions of resonant interaction cease to be fulfilled not only for three waves, but also for four interacting waves. 
In this work, we perform fully nonlinear simulations of plane-symmetric capillary turbulence.
We demonstrate that the system of interacting waves evolves into a quasi-stationary state with a direct energy cascade, despite the absence of low-order resonances. 
The calculated spectra of surface elevations are accurately described by analytical estimates derived dimensionally under the assumption of the dominant influence of five-wave resonant interactions.  A detailed study of the statistical characteristics of the weakly turbulent state does not reveal the influence of any coherent or strongly nonlinear structures. The performed high-order correlation analysis indicates a variety of non-trivial five-wave resonances.  We show that the process of wave decay into two pairs of counter-propagating waves is  responsible for the local energy transfer to small scales. Overall, the calculation results are in good agreement with both the weak turbulence theory and recent experiments made by Ricard and Falcon.

	\end{abstract}
	
	
	\maketitle

\section{ Introduction}

The complex chaotic state of a system of interacting nonlinear waves is called wave turbulence \cite{KZ-book,NazarenkoBook,Newell}. In this regime, energy transfer across scales occurs through resonant wave interactions that conserve both the total frequency and wavevector of the interacting waves. Such resonant interactions are very similar to discrete interactions of  quantum quasiparticles \cite{wave-1,wave-2,wave-3}. In the case of a very slow energy transfer between waves (the time between collisions of quasiparticles is much longer than the wave period), a weakly nonlinear mode of motion can be realized. Such a regime of the system evolution is called Weak Wave Turbulence (WWT). WWT can be described on the basis of a weakly nonlinear expansion of the original equations in a series with respect to a small parameter (wave amplitude). Using the random phase approximation, kinetic equations for the evolution of wave amplitudes are derived. The resulting nonlinear integro-differential equations are extremely complex for analytical investigation. In the late 1960s, Vladimir Zakharov and co-authors first proposed a systematic approach to the analytical solution of these equations \cite{zakh65,zs-70,zf-1,zf-2}. Using a theoretical technique now called the Zakharov transformation, the authors \cite{zakh65,zs-70,zf-1,zf-2} were able to find exact analytical solutions for the kinetic equations describing the steady-state transport of various integrals of motion (energy or number of waves) along scales. This methodology is now known as Weak Turbulence Theory (WTT), and the exact solutions of the kinetic equations are now called Kolmogorov-Zakharov's (KZ) spectra, by analogy with the Kolmogorov spectrum for the classical turbulence.

WTT has been very successful in studying the interaction of nonlinear waves of very different nature. This theory has found a lot of applications in such physical systems as plasma waves,  sea and ocean waves, waves in Bose-Einstein condensates, acoustic and magnetohydrodynamic waves,  and so on, see \cite{KZ-1,plasma-1,plasma-2,Naz22, ShavitFalkovich, FrahmShepelyansky, Semisalov,MHD-1,MHD-2,Galtier1, sound,sound2, sound3,grav-1,grav-2, Galtier, Shuryak,koch-1,koch-2}. Perhaps the most studied type of weak turbulence is the turbulence of dispersive waves on the surface of a liquid, first described in the works of Zakharov and Filonenko \cite{zf-1, zf-2}. To date, the KZ turbulence spectra for gravity and capillary waves (also called the Zakharov-Filonenko spectra) have been confirmed with high accuracy both numerically \cite{PushkarevPRL96, DeikePRL14, PanPRL14,PanJFM15} and experimentally \cite{FalconEPL09,FalconPRL07,KolmakovPLTP09,ARFM2022,CWT23,KZ-2}. The isotropic capillary turbulence has been studied especially well, for which not only the spectra exponents but also the dimensionless coefficient (KZ constant) has been found with high accuracy \cite{KZ-3,KZ-4,cap1,cap2}.

The Zakharov-Filonenko spectrum is usually written in terms of the Fourier spectra of the function $\eta({\bf r},t)$ specifying a two-dimensional surface over a three-dimensional liquid:
\begin{equation}\label{ZF1}
	S_{\eta}(k)=|\eta_{k}|^2=C_{3w}^k P^{1/2}(\sigma/\rho)^{-3/4}k^{-15/4},
\end{equation}
\begin{equation}\label{ZF2}
S_{\eta}(\omega)=|\eta_\omega|^2=C_{3w}^{\omega} P^{1/2}(\sigma/\rho)^{1/6}\omega^{-17/6},
\end{equation}
where $\textbf{k}$ is the wave vector, and $k=|{\bf k}|$ is its absolute value, $\omega$ is the angular frequency, $C_{3w}^k$ and $C_{3w}^{\omega}$ are the dimensionless KZ constants, $P$ is the
energy dissipation rate per unit surface area, $\sigma$ and $\rho$ are the surface tension and mass density of the liquid, respectively. The spatial and frequency spectra given by (\ref{ZF1}) and (\ref{ZF2}) are related to each other by the expression: $S_{\eta}(k)dk=S_{\eta}( \omega)d\omega$, which is the energy conservation law in the Fourier space. The power-law dependencies on $P$ with an exponent of 1/2 in the spectra reflect a resonant character
of three-wave interactions. In general, the conditions for resonant interaction of the $N$-th order can be written as:
\begin{equation}
	\begin{split}
		\omega({\bf k}_1)\pm\omega({\bf k}_2)\ldots\pm\omega({\bf k}_N)&=0, \\
		{\bf k}_1\pm {\bf k}_2\ldots\pm {\bf k}_N&=0. \\
	\end{split}\label{resonance}
\end{equation}
For pure capillary waves, the frequency $\omega$ is related to the wavenumber $k$ by the dispersion relation 
\begin{equation}
\label{DR}
\omega=(\sigma/\rho)^{1/2}k^{3/2}.
\end{equation}
In the case of isotropic surface perturbations, the system (\ref{resonance}) has non-trivial solutions for  $N=3$ that provide a direct energy cascade and, as a consequence, the realization of the Zakharov-Filonenko spectrum (\ref{ZF1}) and (\ref{ZF2}). In a situation of strong anisotropy, when all waves propagate in one direction, i.e., the wave vectors have a single component: ${\bf k}=\{k_x, 0\}$, three-wave resonant interactions degenerate, and the surface evolution begins to be controlled by higher orders of nonlinearity. This situation can be realized in a narrow channel, the width of which is much smaller than the length. Otherwise, the decay instability of a monochromatic wave will inevitably lead to isotropization of turbulence \cite{cap3}. 

Thus, it may seem that a paradoxical situation arises: when moving to a simpler one-dimensional geometry, the behavior of the system of nonlinear capillary waves becomes more complicated. For a long time, the turbulence of capillary waves developing in anisotropic plane-symmetric geometry remained completely unexplored.
For the first time, the turbulence of collinear capillary waves was studied numerically in \cite{koch2020}. The model was based on the equation system, including cubically nonlinear terms. Only trivial three- and four-wave resonant interactions were revealed, which do not lead to the energy transfer over scales. The results of the experimental study \cite{Ricard21} carried out for collinear waves on the surface of mercury indicate that the  resonant five-wave interactions $N = 5$ corresponding to the fourth order of nonlinearity are dominant. At such a high order of nonlinearity, the conditions of resonant interaction (\ref{resonance}) cease to be degenerate. In other words, the system of equations (\ref{resonance}) has exact non-trivial solutions for $N=5$ in the case of plane waves. Having performed a dimensional analysis of weak turbulence spectra \cite{NazarenkoBook}, the authors of \cite{Ricard21} proposed an estimate for the spectrum of capillary wave turbulence in quasi-1D geometry:
\begin{equation}\label{5w1}
	S_{\eta}(k)=C_{5w}^k P^{1/4}(\sigma/\rho)^{-3/8}k^{-27/8},
\end{equation}
\begin{equation}\label{5w2}
S_{\eta}(\omega)=C_{5w}^{\omega} P^{1/4}(\sigma/\rho)^{5/12}\omega^{-31/12},
\end{equation}
where $C_{5w}^k$ and $C_{5w}^\omega$ are the corresponding dimensionless KZ constants. It should be noted that the authors of \cite{five} obtained similar spectra for the plane-symmetric gravity wave turbulence. The experimental results \cite{Ricard21} turned out to be in very good agreement with the formulas (\ref{5w1}) and (\ref{5w2}).

In a recent study \cite{koch23-cwt}, numerical evidence for the realization of the turbulence spectra (\ref{5w1}) and (\ref{5w2}) was obtained within a fully nonlinear plane-symmetric model.
The results of the calculations actually showed that the system of interacting capillary waves can pass into a quasi-stationary turbulent state with a spectrum close to (\ref{5w1}).
It should be noted that in degenerate  1D geometry, so-called coherent structures  (solitons or shock waves) may take a dominant role in the  development of wave turbulence (see, e.g., \cite{rumpf1,rumpf2}). Thus, a convincing demonstration of the dominant influence of resonant five-wave interactions  cannot rely on the quantitative agreement of the results  reported in \cite{koch23-cwt} with spectra (\ref{5w1}) and (\ref{5w2})  only: one needs to  prove that resonances (\ref{resonance}) for $N = 5$ are indeed observed in direct numerical simulations. This is exactly the aim of the present study. In this work, we develop our approach based on the method of dynamic conformal transformations  for the direct numerical simulation of plane-symmetric capillary wave turbulence \cite{koch23-cwt}.
We demonstrate below through the use of high-order correlation functions that the interaction of nonlinear capillary waves is characterized accurately by the conditions (\ref{resonance}). Moreover, in plane-symmetric geometry, a wide class of non-trivial five-wave interactions is found, i.e., not only the processes observed experimentally \cite{Ricard21}, but also many other resonances are realized. In general, the results obtained in this work convincingly demonstrate the realization of a weakly turbulent state of plane-symmetric capillary waves with the KZ spectra (\ref{5w1}) and (\ref{5w2}).

\section{ Numerical Model and Parameters}

The standard way for numerical simulation of WWT is based on the model equations describing weakly nonlinear evolution of a wave system \cite{korot16,Newell,tran23}. In the case of plane-symmetric capillary turbulence, the use of such an approach is complicated: it is necessary to take into account the nonlinear effects of very high (fourth!) order. It is clear that directly solving such a weakly nonlinear system is an extremely difficult task. For this reason, in this work, we use a different approach based on the use of a fully nonlinear equation system describing the evolution of the liquid surface. The current study is based on the method of dynamic conformal transformations, for more details, see \cite{ovs74,dya96, dya2002,tan1,tan2,tan3}. With this approach, the region occupied by the liquid is transformed into a semi-plane of new (conformal) variables. The most important advantage of this method is the reduction in the number of independent variables required to fully describe the system evolution: the final equations are spatially one-dimensional and directly describe the evolution of the surface. The conformal transformation method has proven to be very convenient for describing nonlinear dynamics of liquid surfaces, see, e.g., \cite{ruban20,dya16,korot19,Nachbin,gao19,gao22,koch18,paras1,paras2,paras3}. 
To date, the technique of dynamic conformal transformation has been generalized for a large class of problems describing water waves, including flows with constant vorticity \cite{wwp1,wwp2}, with variable bottom topography \cite{wwp3,wwp4,wwp5}, as well as electrohydrodynamic flows \cite{wwp6,kozu18,kozu14}.
The idea to use the conformal mapping technique for the description of the free-surface wave turbulence was first proposed in our earlier work \cite{koch23-cwt}. Here, we improve the method for detailed simulation of the plane-symmetric capillary turbulence.

This work considers the fully nonlinear dynamics of an ideal incompressible deep liquid with a free surface under the assumption that the motion of the liquid is plane-symmetric; i.e., the full physical model is two-dimensional. Let the Cartesian coordinate system $\{x,y\}$ be such that the equation $y=\eta(x,t)$ specifies the deviation of the free surface from unperturbed
state $y=0$. For the potential of the fluid velocity $\phi(x,y,t)$ the Laplace equation is valid:
\begin{equation}\nonumber
\Delta \phi=0.
\end{equation}
At the liquid surface $y=\eta(x,t)$, the dynamic and kinematic boundary conditions are imposed in the form:
\begin{equation}\label{eqx1}
\phi_t+\frac{1}{2}|\nabla \phi|^2=\frac{\sigma}{\rho}\frac{\eta_{xx}}{(1+\eta_x^2)^{3/2}},
\end{equation}
\begin{equation}\label{eqx2}
\eta_t+\eta_x\phi_x=\phi_y,
\end{equation}
where $\nabla=\{\partial_x, \partial_y\}$. The motion of the liquid vanishes with increasing depth; i.e., $\phi\to 0$ at $y\to -\infty$. The total energy of the system (Hamiltonian) has the form
\begin{equation}\label{Ham}
H=\frac{1}{2}\iint \limits_{y\leq \eta}{|\nabla \phi|^2}dxdy
+\frac{\sigma}{\rho}\int\limits_{-\infty}^{+\infty}\left(\sqrt{1+\eta_x^2}-1\right)dx.
\end{equation}
The system (\ref{eqx1}) and (\ref{eqx2}) can be represented in terms of the variational derivatives of the Hamiltonian (see \cite{zf-1,zf-2}):
\begin{equation} \nonumber
\eta_t=\delta H/\delta \psi,\qquad \psi_t=-\delta H/\delta \eta,
\end{equation}
where the functions $\eta(x,t)$ and $\psi=\phi(x,y=\eta,t)$ are the canonically conjugate variables. The above equation system describes the fully nonlinear evolution of free-surface capillary waves. In this work, we do not take into account the effects of gravity, which corresponds to the consideration of short wavelengths. The transition to dimensionless variables is carried out by choosing some characteristic wavenumber $k_0=2\pi/\lambda_0$, larger than gravity-capillary crossover $k_{gc}=(\rho g/\sigma)^{1/2}$. For air-water interface, $\lambda_{gc}=2\pi/k_{gc}=1.7$~cm. The time scale $t_0$ can be found from (\ref{DR}): $t_0=2\pi/\omega_0=(\lambda_0^3\rho/2\pi\sigma)^{1/2}$. For deep water, the characteristic quantities can be estimated as $\lambda_0\approx1$ cm and $t_0\approx0.05$ s. In the non-dimensional units, 
$$\tilde t=t \omega_0,\quad \tilde x =x k_0,\quad \tilde \eta =\eta/k_0,\quad \tilde \phi =\phi k_0^2/\omega_0,$$	
the dispersion relation (\ref{DR}) takes the form:
\begin{equation}\label{disp}
\omega_k=k^{3/2}.
\end{equation}
For further analysis, the tilde signs are omitted.

We now apply the procedure of conformal transformation of the region occupied by the liquid into the half-plane of new conformal variables: $\{u,v\}$. The surface of fluid corresponds to
the line $v=0$:
\begin{equation}\nonumber
y=Y(u,t),\quad \psi=\Psi(u,t),\quad X=u-\hat H Y(u,t).
\end{equation}
Here, $\hat H$ is the Hilbert transform defined in the Fourier space as $\hat H f_k=i\cdot \mbox{sgn}(k)f_k$. The free surface profile is defined implicitly in parametric form: $\eta(x,t)=Y(X(u,t),t)$. 
To date, the procedure of deriving the equations of motion of the liquid in the conformal variables is well known (see \cite{dya96, dya2002,ovs74}). For this reason, we just write the final equations adding terms responsible for the dissipation and pumping of energy:
\begin{equation}\label{eq1}
Y_t=\left(Y_u\hat H-X_u\right)\frac{\hat H \Psi_u}{J}- \hat \gamma_k Y,
\end{equation}
$$\Psi_t=\frac{(\hat H \Psi_u)^2-\Psi_u^2}{2J}+\hat H\left(\frac{\hat H \Psi_u}{J}\right)\Psi_u+\frac{X_u Y_{uu}-Y_u X_{uu}}{J^{3/2}}$$
\begin{equation}\label{eq2}
+\mathcal{F}(\textbf{k},t)- \hat \gamma_k \Psi,
\end{equation}
where $J=X_u^2+Y_u^2$ is the Jacobian of the transformation, $\hat \gamma_k$ is the dissipation operator, and $\mathcal{F}(\textbf{k},t)$ is the random external force acting at large scales. 
In this work, we take into account the viscosity effect phenomenologically, i.e., high-frequency harmonics are damped starting from a certain wavenumber $k_d$, but in the inertial region of $k$, the dissipation effect is absent. In Fourier space, the viscosity operator $ \hat \gamma _{k}$ is defined as follows:
\begin{eqnarray*}
\hat \gamma _{k} f_k &=&0,\quad k\leq k_{d}, \\ 
\hat \gamma _{k} f_k &=&\gamma _{0} (k-k_d)^4 f_k,\quad k>k_{d}, 
\end{eqnarray*}
where $\gamma _{0}$ is a constant that determines the intensity of dissipation. The fourth power of $k$ in the viscosity operator was taken by analogy with \cite{korot19}. This choice ensures good convergence of the numerical solution.

The term $\mathcal{F}(\textbf{k},t)$ responsible for the energy pumping is defined in Fourier space as:
\begin{eqnarray*}
&\mathcal{F}(\textbf{k},t)& =F(k)\cdot \exp [iR(\textbf{k},t)], \\
&F(k)& =F_{0}\cdot \exp [-L_0^4(k-k_{f})^{4}].
\end{eqnarray*}
Here, $R(\mathbf{k,}t)$ are random numbers uniformly distributed in the interval $[0,2\pi ]$,  $F_{0}$ is a constant, $k_f$ is the wavenumber at which the pumping amplitude reaches the maximum, and $L_0$ specifies the width of pumping.  In terms of the conformal variables, the total energy of the system (\ref{Ham}) is represented in the form:
\begin{equation}\label{ham2}
H=\frac{1}{2}\int\limits_{-\infty}^{+\infty} \left[ -\Psi \hat H \Psi_u+2(J^{1/2}-X_u)\right]du,
\end{equation}
where the first term in the integrand corresponds to the kinetic energy of the system and the second term is the potential energy of surface capillary waves. In the
case of an infinitesimally small amplitude of surface waves and zero pumping and dissipation of energy, equations (\ref{eq1}) and (\ref{eq2}) are transformed into the dispersion relation (\ref{disp}). In the linear approximation, the key quantities take the form: $ u\to x$, $Y\to \eta(x,t)$ and $\Psi\to \psi(x,t)$.

The purpose of this work is to accurately numerically solve the system of nonlinear integro-differential equations (\ref{eq1}) and (\ref{eq2}). To do this, we use pseudo-spectral methods based on the fast Fourier transform, as a result, the boundary conditions are periodic in space.
The total number of Fourier harmonics used is equal to $N_T$. The time integration is performed by an explicit fourth-order Runge–Kutta method with the step $dt$. All calculations are carried out in a periodic region with the length $L=2\pi$ with the following parameters: $dt=2.5\cdot 10^{-6}$, $N_{T}=8192$, $\gamma_0=10^{ -6}$, $k_d=750$, $F_0=1.5\cdot 10^5$, $k_f=3$, $L_0=0.64$. To suppress the aliasing effect, a low-frequency filter eliminating higher harmonics $k\geq N_T/3$ is used at each step of the integration in time.

\section{ Simulation results} 

This Section presents the results of numerical simulations of capillary wave turbulence based on the system of equations (\ref{eq1}) and (\ref{eq2}) using the selected parameters. The total simulation time was set to $t=500$, corresponding to approximately 100 characteristic periods $t_0$ of the slowest large-scale harmonic (with wavenumber $k=k_0$).

\subsection{Quasi-stationary chaotic state}

\begin{figure}[t]
	\centering
	\includegraphics[width=1.0\linewidth]{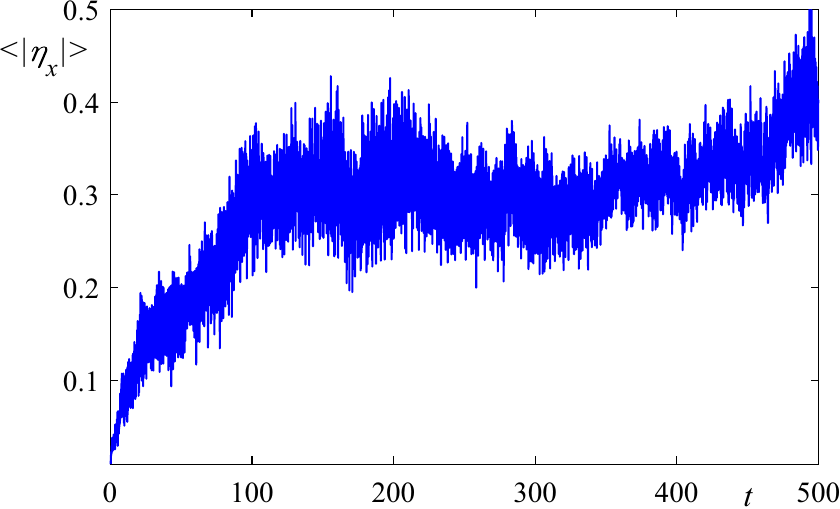}
	\caption{ Time evolution of the spatially averaged wave steepness.}
	\label{fig1}
\end{figure}

\begin{figure}[t]
	\centering
	\includegraphics[width=1.0\linewidth]{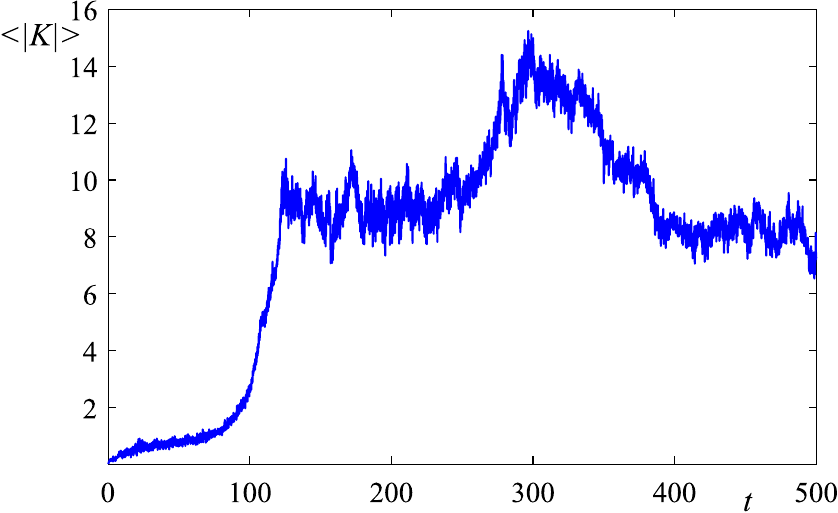}
	\caption{ Time evolution of the spatially averaged wave curvature.}
	\label{fig2}
\end{figure}

First, we  show that under the influence of a randomly driving force acting on large scales and small-scale dissipation, a stationary chaotic state (wave turbulence) arises.
Figure~\ref{fig1} shows the evolution of the spatially averaged wave steepness: $\langle|\eta_x|\rangle=L^{-1}\int |\partial \eta/\partial x|dx$. It can be seen that the average steepness of the waves increases at the beginning of the calculation interval, and by the time  $t\approx 100$ its growth stabilizes. Then the steepness of the waves oscillates around the average value, defined as:
\begin{equation}\label{steep}\epsilon=\left\langle \sqrt{\frac{1}{L}\int|\partial \eta/\partial x|^2dx}\right\rangle_t \simeq 0.3,
\end{equation}
which is noticeably smaller than unity. 

Figure~\ref{fig2} shows the average value of the surface curvature $K$ defined as follows:
\begin{equation*}\langle |K|\rangle=L^{-1}\int\left|\frac{\eta_{xx}}{(1+\eta_x^2)^{3/2}}\right|dx,
\end{equation*}
which is responsible for the influence of surface tension effects. One can see that the temporal behavior of the average curvature is qualitatively the same as the evolution of the wave steepness. Both quantities increase in the early stages of evolution, but then their growth stops due to dissipative effects.  Thus, we actually observe a transition to a quasi-stationary state, in which the energy pumping effect is completely stabilized by dissipation. Note that the total energy also reaches saturation in the quasi-stationary state. The average energy (\ref{ham2}) in the steady state is near the value $\langle H \rangle \approx 1$.

\begin{figure}[t]
	\centering
	\includegraphics[width=1.0\linewidth]{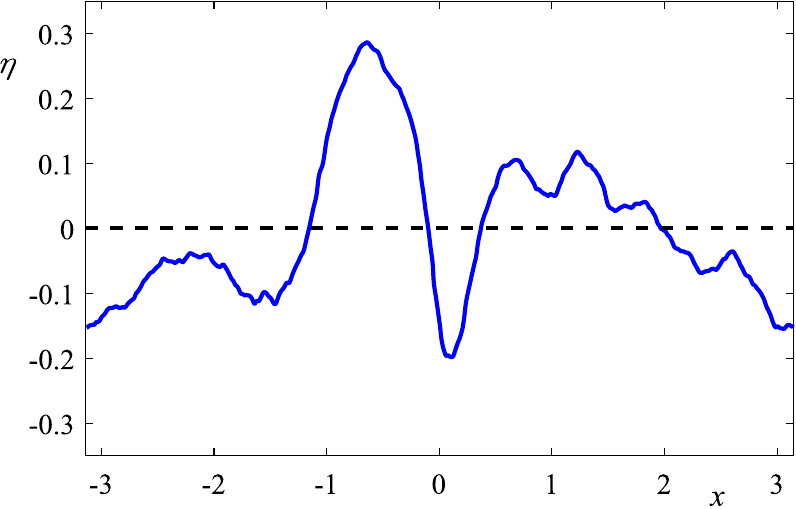}
	\caption{Shape of the liquid surface is shown in the quasi-stationary state at the time moment $t=450$.}
	\label{fig3}
\end{figure}

\begin{figure}[t]
	\centering
	\includegraphics[width=1.0\linewidth]{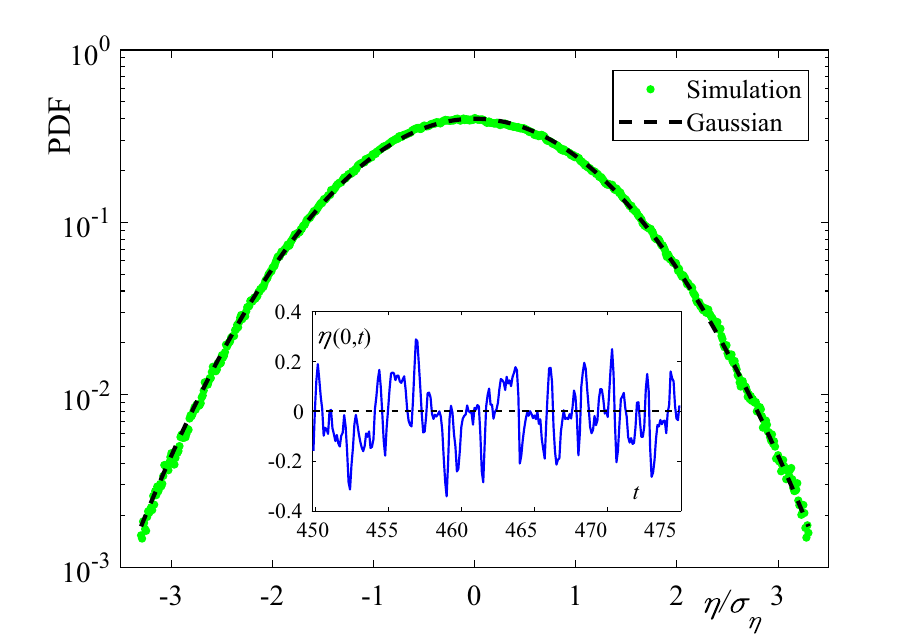}
	\caption{The probability density function (PDF) for surface amplitudes measured with respect to the standard deviation $\sigma_{\eta}$ in comparison with the Gaussian distribution presented by the black dashed line. The inset shows the evolution of surface elevation at a point $x=0$ over the time interval $t \in [450,475]$.}
	\label{fig4}
\end{figure}

In the quasi-stationary state, the dynamics of the liquid surface becomes complex and chaotic, see Figure~\ref{fig3}, which demonstrates the boundary shape at the time moment $t=450$. Figure~\ref{fig4} shows the probability density function (PDF) for the surface amplitude measured in the steady state. One can see that the probability density function is very close to the Gaussian distribution shown by the black dashed line. The influence of any coherent structures that manifest themselves in strong non-Gaussianity (intermittency) of the PDF is not detected. Thus, strongly nonlinear structures such as jets, droplets, and bubbles  \cite{strong} are not observed on the liquid surface at the steepness of (\ref{steep}).

\subsection{Turbulence spectra}

The main question of the present study is whether the turbulence spectra (\ref{5w1}) and (\ref{5w2}) are indeed observed in plane-symmetric capillary wave turbulence. To answer the question, it is necessary to restore the explicit dependency: $y=\eta(x,t)$, since the function $Y(u,t)$ calculated directly from the equations (\ref{eq1}) and (\ref{eq2}) is defined not in physical space but in the auxiliary conformal coordinate $u$. The parametric dependence $y=Y(X_i)$ obtained from the equations (\ref{eq1}) and (\ref{eq2}) is defined at points $i$ of the calculated nonuniform grid $X_i(u,t)$. To obtain the explicit dependence $y=\eta(x,t)$, it is necessary to interpolate the values of the function $Y(X(u))$ at new points $j$ with a fixed step $dx=X_{j+1}-X_{j}$. The procedure for such interpolation was proposed in our earlier work \cite{koch23-cwt} and is based on using the cubic splines. This interpolation allows us to obtain an explicit dependence $y=\eta(x,t)$ with high accuracy in the case of a sufficiently dense grid and the absence of strongly nonlinear structures in which the function $Y(X(u))$ loses to be single valued.
When these conditions are met, the relative error in determining the energy (\ref{ham2}) does not exceed a small value of $10^{-4}$.

\begin{figure}[t]
	\centering
	\includegraphics[width=1.0\linewidth]{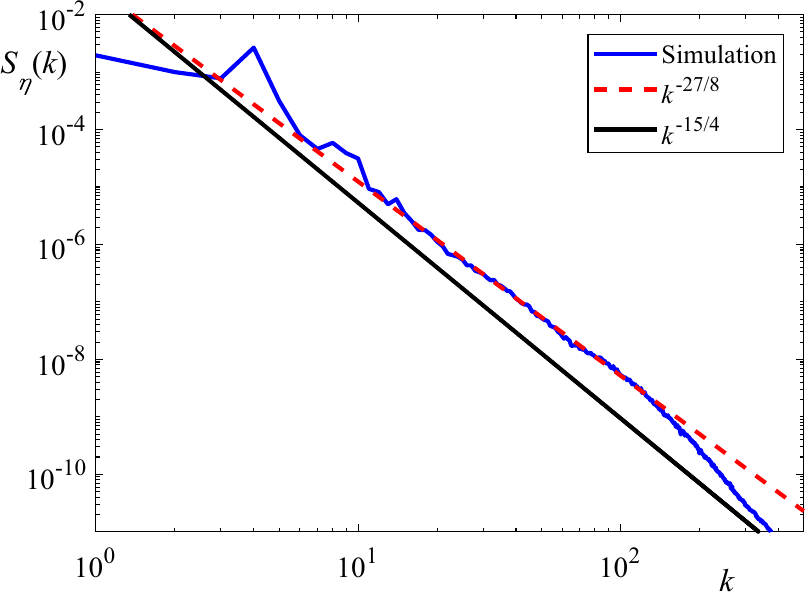}
	\caption{The spatial spectrum of surface elevations $S_{\eta}(k)$ averaged over time in the quasi-stationary state. The black solid line is the Zakharov–Filonenko spectrum (\ref{ZF1}), and the red dashed line is the turbulence spectrum (\ref{5w1}) arising due to five-wave resonant interactions.  }
	\label{fig5}
\end{figure}

\begin{figure}[t]
	\centering
	\includegraphics[width=1.0\linewidth]{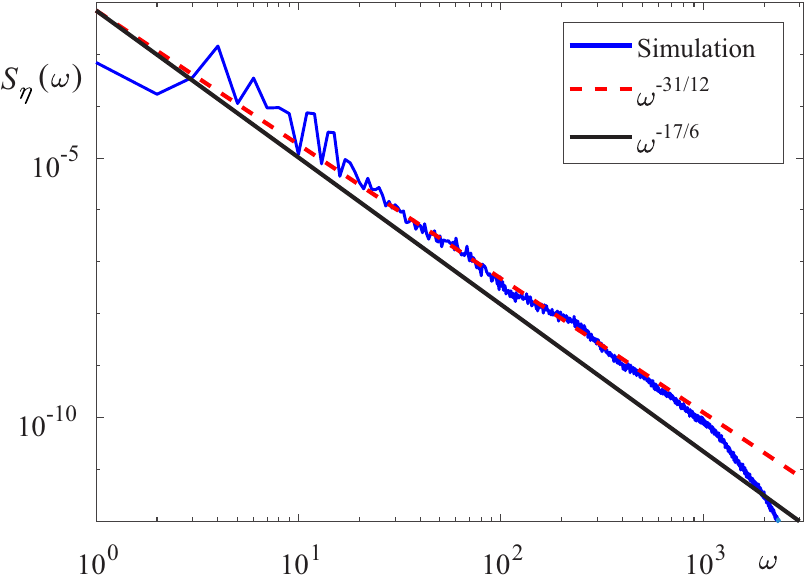}
	\caption{The frequency spectrum of surface elevations $S_{\eta}(\omega)$ averaged over space interval $[-\pi, \pi]$ in the quasi-stationary state. The black solid line is the Zakharov–Filonenko spectrum (\ref{ZF2}), and the red dashed line is the turbulence spectrum (\ref{5w2}) arising due to five-wave resonant interactions. }
	\label{fig6}
\end{figure}

Figure~\ref{fig5} shows the spatial spectrum of the restored shape of the surface $S_{\eta}(k)=\left\langle |\eta_k|^2\right\rangle_t$ averaged over time in the quasi-stationary  regime of motion. One can see that the spectrum of surface perturbations is well described by a power-law in the wavenumber interval $k\in[10, 150]$, which is more than one decade.
The spectrum exponent differs significantly from the classical Zakharov–Filonenko spectrum (\ref{ZF1}), obtained under the assumption of the dominant role of three-wave resonant interactions in isotropic capillary turbulence.  At the same time, the simulation results are much better described by the analytical spectrum (\ref{5w1}) derived under the assumption of the dominant role of five-wave resonant interactions. It is also important to compare the calculated frequency spectrum in the developed wave turbulence regime with (\ref{5w2}). To do this, we performed a temporal Fourier transform of the surface shape. The calculated frequency spectrum $S_{\eta}(\omega)$ is shown in Figure~\ref{fig6}. Here we see the same behavior as in  Figure~\ref{fig5}: the analytical estimate (\ref{5w2}) is closer to the calculated turbulence spectrum than the Zakharov-Filonenko spectrum (\ref{ZF2}). Thus, the obtained turbulence spectra demonstrate good agreement with both the theoretical predictions of WTT and the experimental results reported by Ricard and Falcon \cite{Ricard21}.

\subsection{Nonlinear dispersion relation}

\begin{figure}[t]
	\centering
	\includegraphics[width=1.0\linewidth]{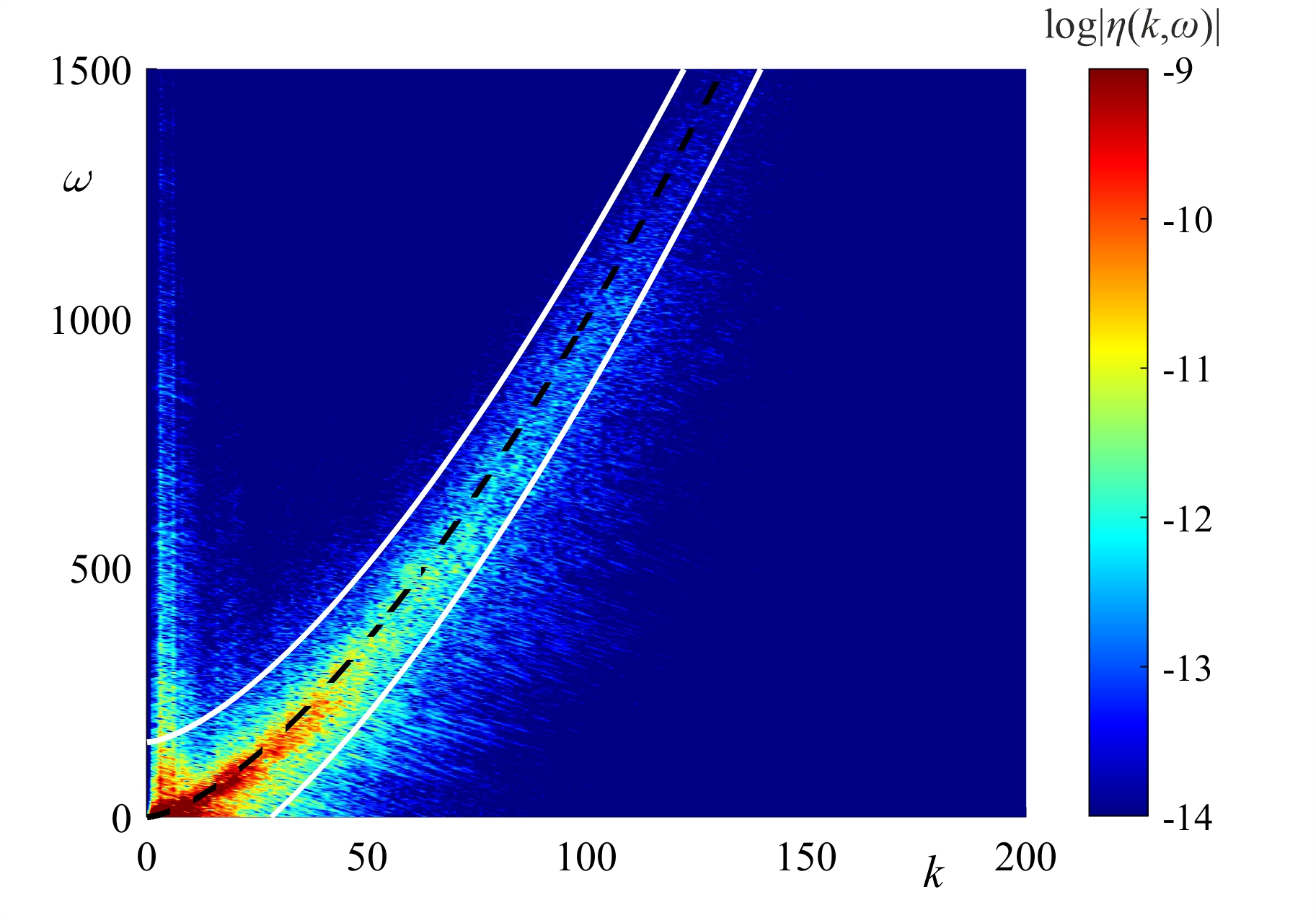}
	\caption{The space-time Fourier transform $|\eta(k,\omega)|$ is presented in logarithmic scale. The black dashed line corresponds to the exact dispersion relation given by (\ref{disp}), and the white solid lines describe the nonlinear frequency broadening: $\omega_k\pm\delta_{\omega}$ with $\delta_{\omega}=50$. }
	\label{fig7}
\end{figure}

\begin{figure}[t]
	\centering
	\includegraphics[width=1.0\linewidth]{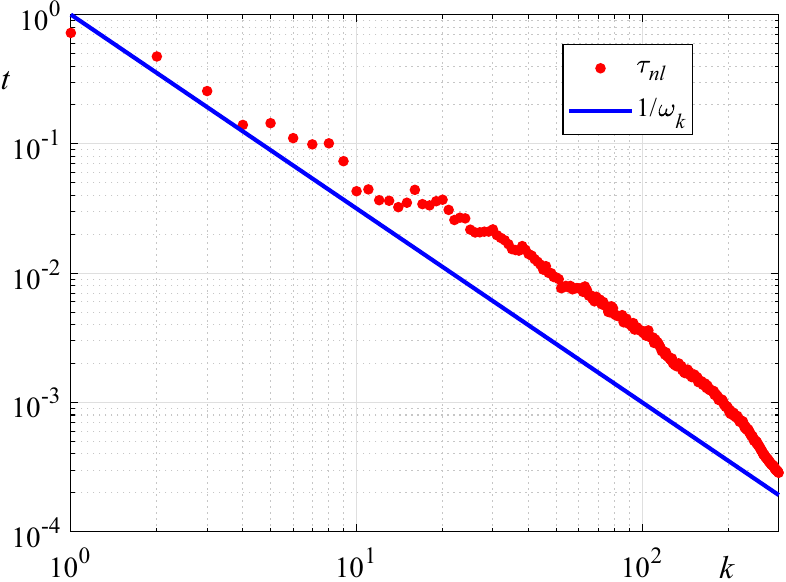}
	\caption{Characteristic nonlinear time $\tau_{nl}$ measured based on the space-time Fourier transform presented in Figure~\ref{fig7}. The blue solid line corresponds to the linear time determined from the dispersion law (\ref{disp}).}
	\label{fig8}
\end{figure}

It is known that in low-dimensional nonlinear wave systems, e.g., the one-dimensional MMT model \cite{MMT1,MMT2,MMT3}, strongly nonlinear and coherent structures can significantly influence the mechanism of energy transfer along scales. At a sufficiently high level of nonlinearity, so-called bound waves can arise in gravity wave turbulence \cite{bound}. Such waves are generated as a result of non-resonant wave interactions due to small but finite nonlinearity. In the situation under study, the level of nonlinearity  is relatively high (mean steepness is near 0.3), and one can expect that various coherent structures can influence the evolution of plane capillary waves.
To study the possible effects of coherent structures on the free surface evolution, we present the space–time Fourier transform of the boundary shape in Figure~\ref{fig7}. 
It is evident from the figure that perturbations are found in a wide range of wavenumbers. Fourier harmonics are distributed in a narrow region along the linear dispersion relation (\ref{disp}). 
Coherent and strongly nonlinear structures that would significantly distort the dispersion law are not detected in the system under study. Figure~\ref{fig7} shows only a simple frequency broadening due to nonlinearity (dependence of wave speed on amplitude).  The vertical band in the region of small $k$ is associated with energy pumping since the external random force is not localized in the region of low frequencies.

The average broadening of the frequency $\delta_{\omega}(k)$ for a given wavenumber $k$ can be estimated by the formula \cite{Naz22}:
\begin{equation}\label{time}
\delta_{\omega}(k)=\left[\frac{\int_0^{\infty}(\omega-\omega_k)^2|\eta(k,\omega)|^2 d\omega}{\int_0^{\infty}|\eta(k,\omega)|^2 d\omega}\right]^{1/2},
\end{equation}
where $\omega_k$ is determined from the dispersion relation (\ref{disp}). The quantity $\delta_{\omega}$ characterizes the nonlinear time $\tau_{nl}=1/\delta_{\omega}$, which is an important parameter in WTT \cite{KZ-book,NazarenkoBook}. 
The condition of smallness of nonlinear effects can be formulated as: $\tau_{nl}\gg\tau_{l}$, where $\tau_{l}=1/\omega_k$ is the time determined from the linear dispersion law (\ref{disp}). From a physical point of view, such a condition means that the probability of interaction of waves is small during their wave period, and the dimensionless frequency of collisions can be expressed as: $\tau_l/\tau_{nl}\ll 1$. Figure~\ref{fig8} shows the nonlinear time $\tau_{nl}$ calculated using the formula (\ref{time}) compared to the linear one $\tau_{l}$. One can indeed see that the measured quantity $\tau_{nl}$ is indeed much greater than linear time $\tau_l$ over the entire range of wavenumbers.
	Thus, the comparison of characteristic times undoubtedly indicates a weakly nonlinear nature of the system evolution. It should be noted that when the nonlinearity level exceeds a certain threshold value, weak turbulence can transform into a strongly nonlinear state in which various coherent structures (for example, solitons, shock waves, etc.) dominate \cite{Kuz-2004,koch-25,wwp7,wwp8,wwp9,wwp10,wwp11}.

\subsection{Estimation of the Kolmogorov-Zakharov constant}

\begin{figure}[t]
	\centering
	\includegraphics[width=1.0\linewidth]{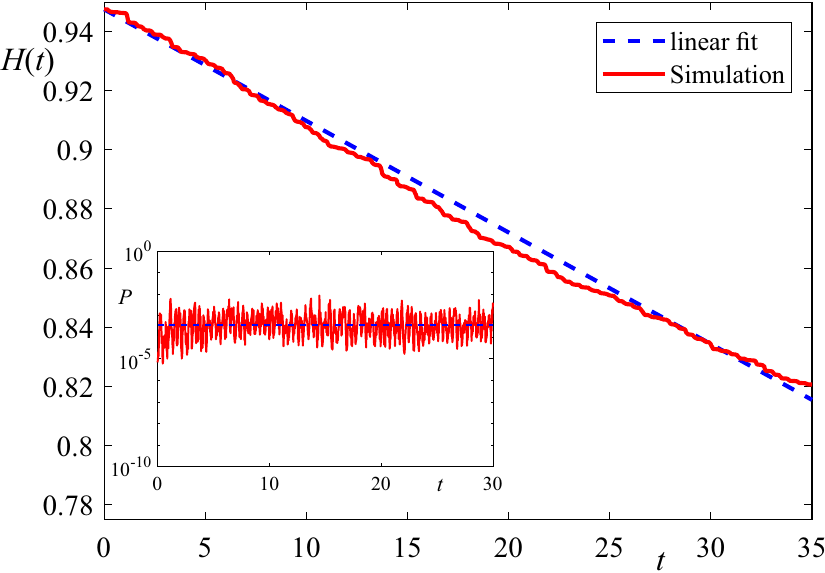}
	\caption{Evolution of the total energy of the system in the regime of freely decaying turbulence. The blue dashed line describes decay at a constant rate. The inset shows the corresponding energy dissipation flux $P$.}
	\label{fig9}
\end{figure}

\begin{figure}[t]
	\centering
	\includegraphics[width=1.0\linewidth]{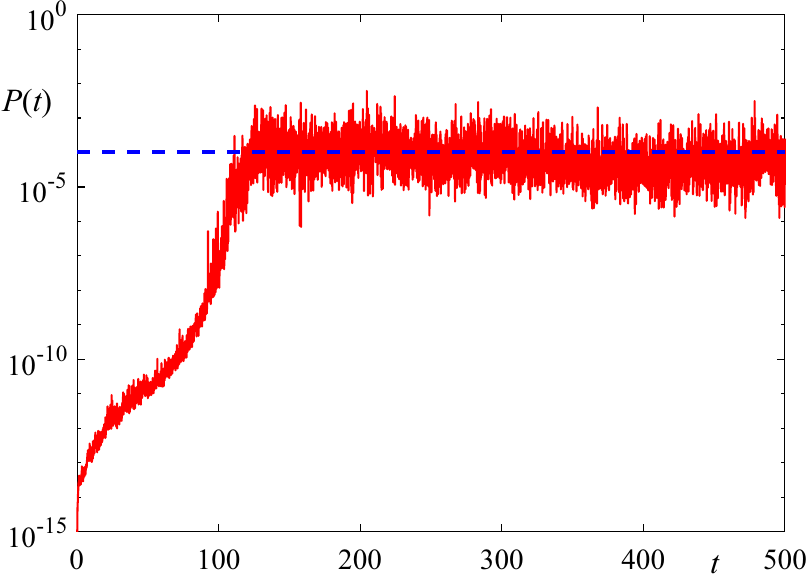}
	\caption{The calculated instantaneous energy dissipation flux $P(t)$ is shown as a function of time in the regime of forced turbulence. The blue dashed line shows the averaged value of the flux at the steady state, $\langle P\rangle =10^{-4}$. }
	\label{fig10}
\end{figure}

An important characteristic of the weakly turbulent state of a wave system is the dimensionless KZ constant, which is a part of the turbulence spectra (\ref{5w1}) and (\ref{5w2}). By definition, the quantities $C_{5w}^k$ and $C_{5w}^{\omega}$ are related to each other as: $C_{5w}^k/C_{5w}^{\omega}=3/2$. The comprehensive research conducted by Ricard and Falcon \cite{Ricard21} provided an experimental estimate of the KZ constant $C_{5w}^k\approx5.4 \cdot 10^{-3}$. In order to estimate the KZ constant numerically, we will use two approaches. The first one is based on the simulation of freely decaying turbulence. Figure~\ref{fig9} shows the time dependence of the total energy of the system (\ref{ham2}) in the regime of decaying turbulence. To obtain the data, we set $F_0=0$, and the initial conditions are taken from the final moment of time $t=500$. It is clearly seen that the energy decays at an almost constant rate. This is due to the selected time interval is quite short. At longer times, the energy will decay exponentially but not linearly.
The inset to Figure~\ref{fig9} shows the corresponding energy dissipation flux, which we define as follows: $P(t)={(2\pi)}^{-1}|dH(t)/dt|$. The average energy flux reaches the value $\langle P \rangle\approx 5.7 \cdot 10^{-4}$. Finally, using the data from  Figures~\ref{fig5}~and~\ref{fig6}, one can obtain estimates for the corresponding KZ constants for plane-symmetric capillary turbulence: $C_{5w}^k\approx0.19$ and $C_{5w}^{\omega}\approx0.04$. The ratio of the calculated constants $C_{5w}^k/C_{5w}^{\omega}\approx 4$ differs from the theoretical value $3/2$, but the deviation does not exceed an order of magnitude.

The second, more accurate approach for estimating the Kolmogorov-Zakharov constant involves evaluating the instantaneous energy dissipation flux in the stationary turbulence regime. Following the approach proposed in \cite{PanPRL14}, this method can be readily applied within the framework of equations (\ref{eq1}) and (\ref{eq2}). The instantaneous dissipation flux is computed as follows:
\begin{equation}\label{flux1}
	P(t)=\frac{1}{2\pi}\int \limits_{k > k_d} \hat \gamma_k \left(kS_{\Psi}(k)+k^2S_{Y}(k)\right)dk,
	\end{equation}
where $\hat \gamma_k $ is the viscosity operator, $S_{\Psi}(k)=|\Psi_k|^2$ and $S_{Y}(k)=|Y_k|^2$ are the Fourier spectra of the functions $\Psi$ and $Y$. A direct calculation of the flux (\ref{flux1}) over the entire time interval is shown in  Figure~\ref{fig10}. The figure correlates very well with Figures~\ref{fig1}~and~\ref{fig2}. The quantity $P(t)$ increases rapidly and reaches a stationary value close to $10^{-4}$ by the time $t\approx 100$. Note that our estimate for the flux based on decaying turbulence shown in the inset in Figure~\ref{fig9} is close to the value $\langle P\rangle =10^{-4}$ obtained in the steady-state turbulence regime. The corresponding estimate for the KZ constant is $C_{5w}^k\approx 0.29$, which is also close to the previously obtained value 0.19.

Thus, the value of the KZ constant averaged over two methods is:
\begin{equation}C_{5w}^k\approx0.24\pm0.05, \label{KZ}
	\end{equation}
which differs significantly from the experimental estimate $C_{5w}^k\approx5.4 \cdot 10^{-3}$. The difference is two orders of magnitude. Within the framework of this work, it is very difficult to explain such a significant distinction. Perhaps the disagreement is explained by different estimation methods, another level of nonlinearity, or the influence of gravity, which cannot be fully suppressed experimentally. The calculated value of $C_{5w}^k$ is much closer to the theoretical estimates of the KZ constant for isotropic capillary turbulence  $C_{3w}^k=9.85$ by Pushkarev and Zakharov \cite{KZ-3}, and  $C_{3w}^k=6.97$ (more accurate estimation by Pan and Yue \cite{KZ-4}), for which three-wave resonant interactions dominate.

\section{High-order correlation analysis }
The question of the order of resonant interaction within the system of equations (\ref{resonance}) is a key issue in constructing the weak turbulence theory. To answer it, in this part of the work, we apply high-order correlation functions widely used in statistical and nonlinear physics \cite{cor1,cor2,cor3,cor4}.

\subsection{Bicoherence}

First, we consider the normalized third-order correlation function (bicoherence), describing three-wave interactions:
\begin{equation}
	B(k_1,k_2)=\frac{|\langle \eta_{k_1}\eta_{k_2}\eta_{k_1+k_2}^*\rangle|}{\sqrt{\langle|\eta_{k_1}\eta_{k_2}|^2\rangle\langle|\eta_{k_1+k_2}|^2\rangle}}  {\rm \ ,}
	\label{bic}
\end{equation}
where, $\eta_k$ is the spatial Fourier transform of the function $\eta(x,t)$ calculated at time instant $t$,  $\eta_k^\ast$ is complex conjugate of  $\eta_k$, and angle brackets denote time averaging. To compute the correlators, the wave evolution was analyzed over a short time interval of the stationary state with a duration of 1 and a temporal resolution of 0.01. Complete correlations correspond to $B(k_1,k_2)=1$ and absence of any correlations is $B(k_1,k_2)=0$. The correlator (\ref{bic}) describes the processes of decay of a wave with the wavenumber $k_3=k_1+k_2$ into two others with wavenumbers $k_1$ and $k_2$ or, in other words, the process $1\to 2$.

\begin{figure}[t]
	\centering
	\includegraphics[width=1.0\linewidth]{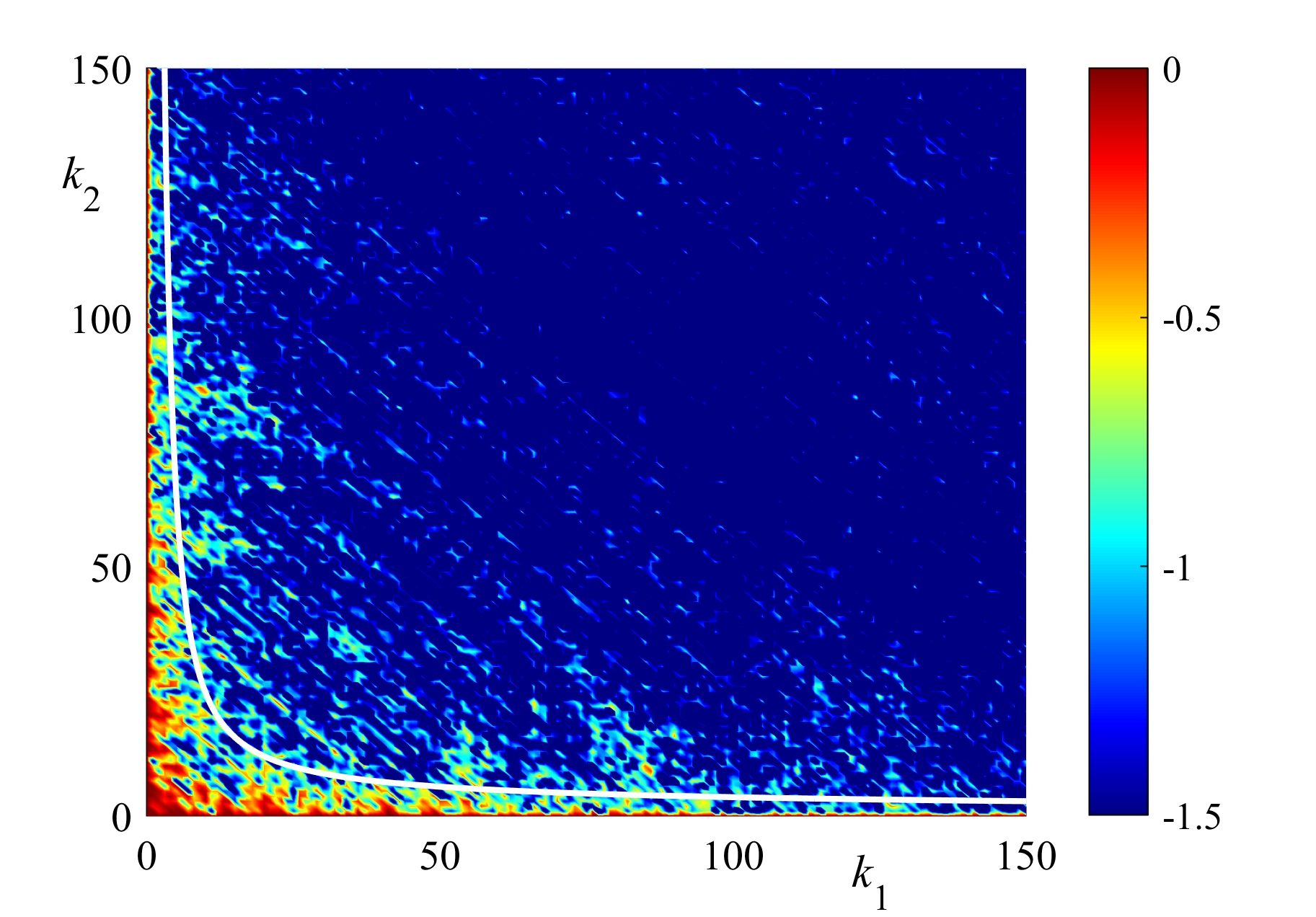}
	\caption{Bicoherence $B(k_1,k_2)$ calculated in the quasi-stationary state is shown in logarithmic scale. The white solid line corresponds to the border of the quasi-resonant interaction area defined from (\ref{reson2}) with $\delta_{\omega}=50$.}
	\label{fig11}
\end{figure}

 \begin{figure*}[t]
	\centering
	\includegraphics[width=0.85\linewidth]{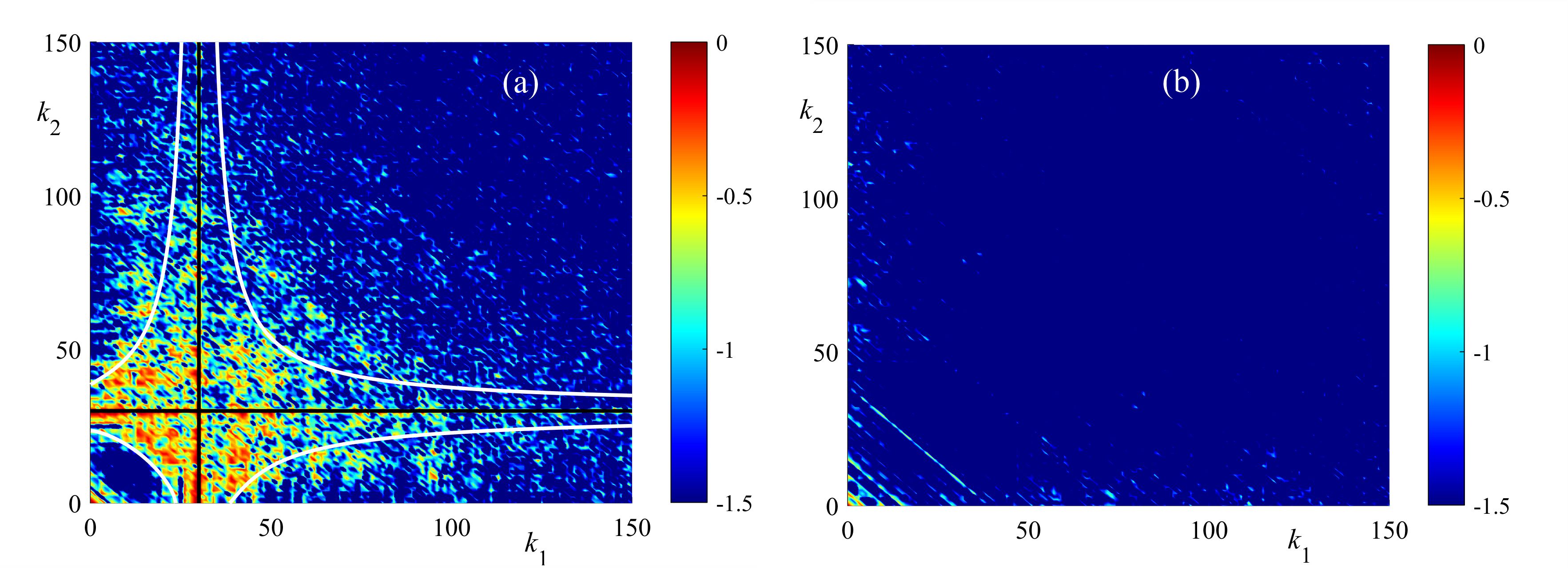}
	\caption{Logarithm of tricoherence $T(k_1,k_2,k_3)$ (\ref{tric1}) computed for the $2\to2$ processes (a),  and (b) is the correlator (\ref{tric2}) responsible for the wave disintegration $1\to3$. Black solid lines show the trivial solutions of the system (\ref{resonance}) and the white curves  correspond to the border of the quasi-resonant interaction area defined from (\ref{reson2}) with $\delta_{\omega}=50$. The wavenumber $k_3$ is fixed, $k_3=30$.}
	\label{fig12}
\end{figure*}

The calculated bicoherence is shown in Figure~\ref{fig11}. It can be seen that correlations are only detected near the coordinate axes. Analytically, the system (\ref{resonance}) has no solutions for $N=3$ in plane-symmetric geometry. The correlations visible in Figure~\ref{fig11} arise as a result of the so-called quasi-resonant interactions, when instead of an equation for frequencies in (\ref{resonance}), an inequality arises:
\begin{equation}
	\begin{split}
		\omega({ k}_1)\pm\omega({ k}_2)\ldots\pm\omega({ k}_N)&\leq \delta_{\omega}, \\
		{ k}_1\pm { k}_2\ldots\pm { k}_N&=0. \\
	\end{split}\label{reson2}
\end{equation}
Such quasi-resonant interactions arise as a result of nonlinear frequency broadening observed in Figure~\ref{fig7}. The calculation results are in complete agreement with the conditions (\ref{resonance}) and (\ref{reson2}): no three-wave resonances are found in plane-symmetric geometry.

\subsection{Tricoherence}

The next third order of nonlinearity is described by a fourth-order correlation function called tricoherence $T(k_1,k_2,k_3)$. In contrast to the simplest three-wave processes, which represent the decay of waves $1\to2$, four-wave resonances include interactions of two types: $2\to2$ and $1\to3$.  The correlator responsible for the processes $2\to2$ is written as:
\begin{equation}
	T(k_1,k_2,k_3)=\frac{|\langle\eta_{k_1}\eta_{k_2}\eta^*_{k_3}\eta^*_{k_1+ k_2- k_3}\rangle|}{\sqrt{\langle|\eta_{k_1}\eta_{k_2}|^2\rangle\langle|\eta_{k_3}\eta_{k_1+ k_2 - k_3}|^2\rangle}},
	\label{tric1}
\end{equation}
which corresponds to the condition:
\begin{equation}\label{4w1}
k_1+ k_2=k_3+ k_4.
\end{equation}
The processes of 1 $\to$ 3 ($k_4=k_1+ k_2+k_3$) can be detected by the following form of tricoherence:
\begin{equation}
T(k_1,k_2,k_3)=\frac{|\langle\eta_{k_1}\eta_{k_2}\eta_{k_3}\eta^*_{k_1+ k_2+ k_3}\rangle|}{\sqrt{\langle|\eta_{k_1}\eta_{k_2}\eta_{k_3}|^2\rangle\langle|\eta_{k_1+ k_2 + k_3}|^2\rangle}}.
\label{tric2}
\end{equation}

Figure~\ref{fig12} shows the computed fourth-order correlators for the possible four-wave processes. To plot the tricoherence on a two-dimensional graph, we fixed one of the wavenumbers: $k_3=30$.
For the processes $2\to2$ shown in Figure~\ref{fig12}~(a),  the calculated correlations lie along straight lines. These lines correspond to the trivial solutions of the system (\ref{resonance}): $k_1=k_3$ and $k_2=k_3$ for the process (\ref{4w1}). The figure also shows a huge number of quasi-resonant interactions, the boundaries of which are indicated by white solid lines. It is precisely such interactions that were first described based on the cubically nonlinear model \cite{koch2020}. The observed trivial or, as they say, degenerate resonances do not lead to energy transfer to small scales \cite{NazarenkoBook}. At the same time, the processes of decay of one wave into three co-directed waves $1\to 3$, described by the correlator (\ref{tric2}), are prohibited in plane-symmetric geometry. The measured tricoherence (\ref{tric2}) is shown in Figure~\ref{fig12}~(b). In the figure, we do not observe any correlations nor any exact solutions  of the system (\ref{resonance}). In general, the analysis carried out leads to the need to consider resonances of a higher order.

\subsection{Quadricoherence}

\begin{figure*}[t!]
	\centering
	\includegraphics[width=1.0\linewidth]{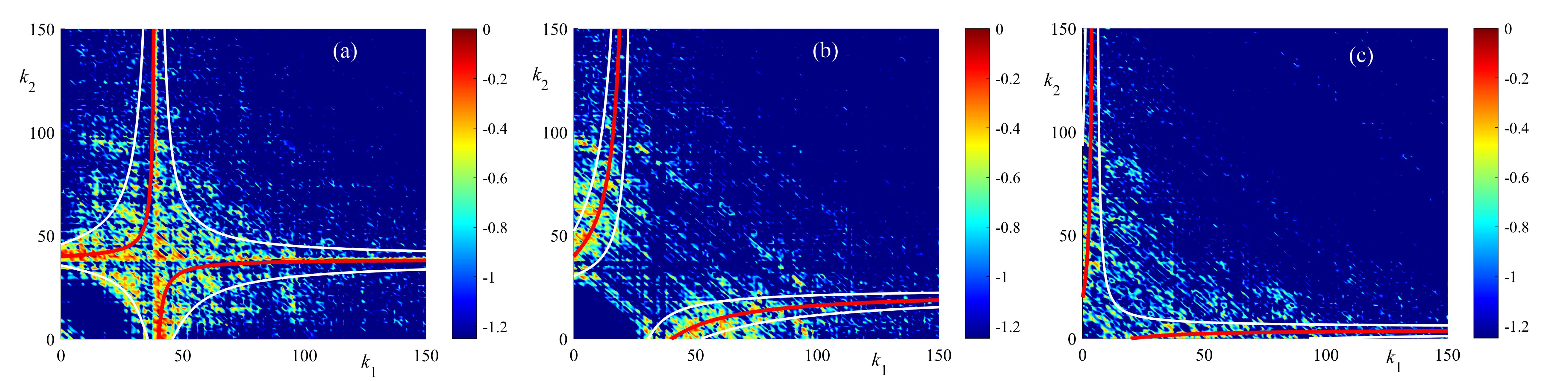}
	\caption{Logarithm of quadricoherence $Q(k_1,k_2,k_3,k_4)$ (\ref{quadr1}) for the process  (\ref{3to2}): (a) $k_3=1$, $k_4=40$, (b) $k_3=15$, $k_4=40$,  and (c) $k_3=15$, $k_4=20$. Red solid lines show the non-trivial solutions of the system (\ref{resonance}) for $N=5$, and the white curves  correspond to the border of the quasi-resonant interaction area defined from (\ref{reson2}) with $\delta_{\omega}=50$.}
	\label{fig13}
\end{figure*}

\begin{figure*}[t]
	\centering
	\includegraphics[width=0.85\linewidth]{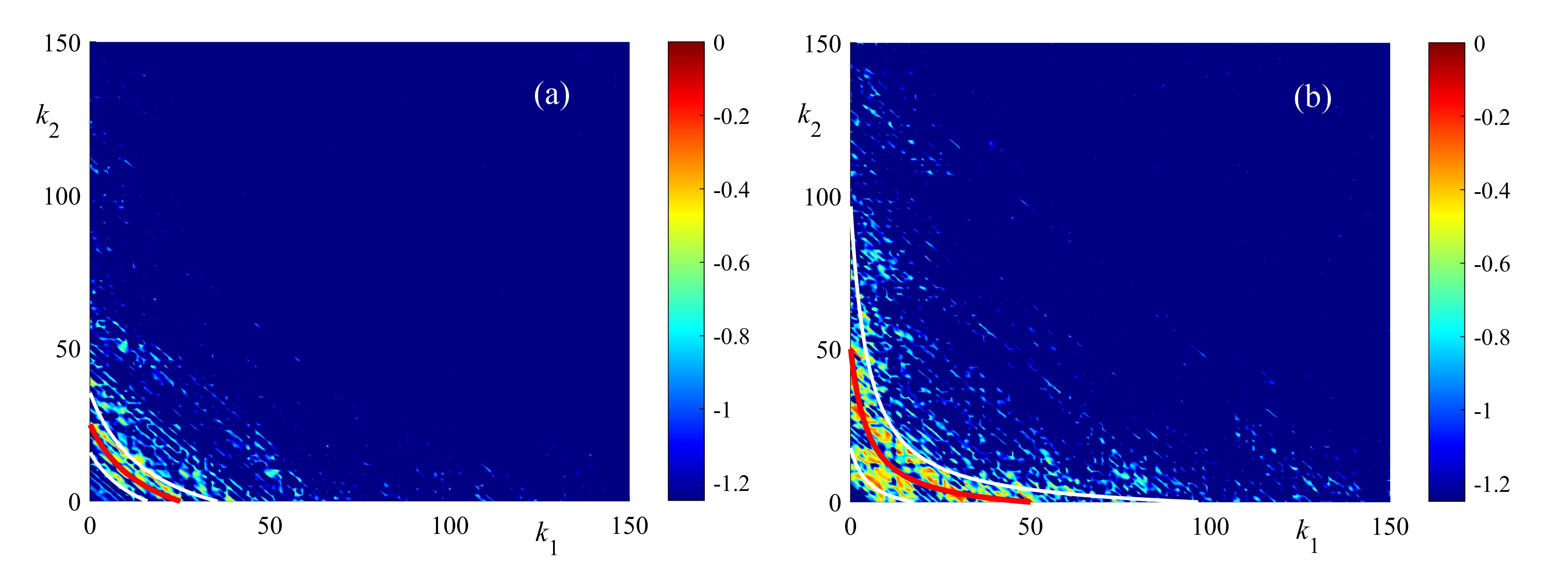}
	\caption{Logarithm of quadricoherence $Q(k_1,k_2,k_3,k_4)$ (\ref{quadr1}) for the process  (\ref{3to2}): (a) $k_3=60$, $k_4=25$, (b) $k_3=60$, $k_4=50$. Red solid lines show the non-trivial solutions of the system (\ref{resonance}) for $N=5$, and the white curves correspond to the border of the quasi-resonant interaction area defined from (\ref{reson2}) with $\delta_{\omega}=50$.}
	\label{fig14}
\end{figure*}

\begin{figure*}[t]
	\centering
	\includegraphics[width=1.00\linewidth]{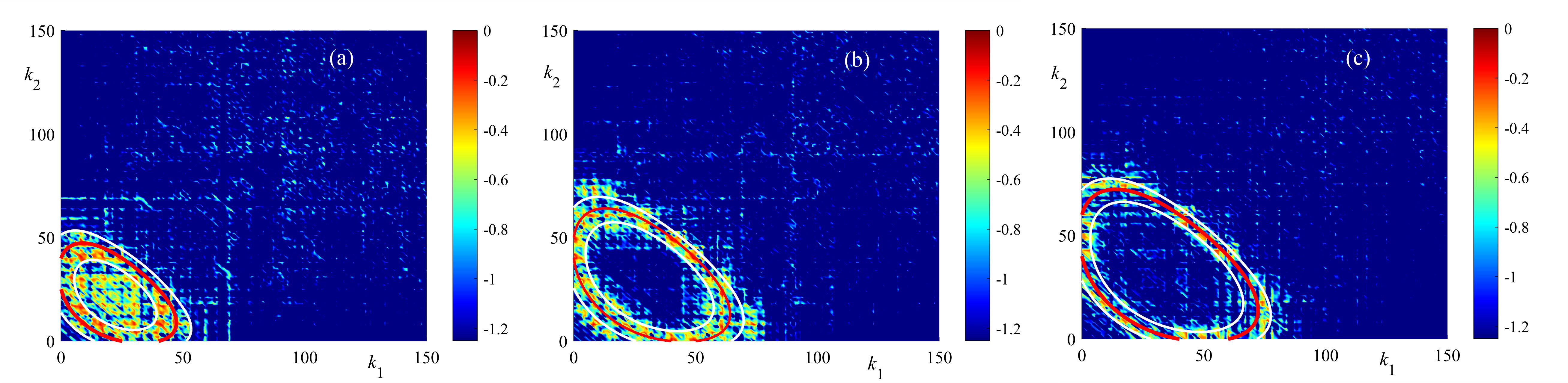}
	\caption{Logarithm of quadricoherence $Q(k_1,k_2,k_3,k_4)$ (\ref{quadr3}) for the process  (\ref{2to3}): (a) $k_3=25$, $k_4=40$, (b) $k_3=50$, $k_4=40$, and (c) $k_3=60$, $k_4=40$. Red solid lines show the non-trivial solutions of the system (\ref{resonance}) with (\ref{2to3n}), and the white curves correspond to the border of the quasi-resonant interaction area defined from (\ref{reson2}) with $\delta_{\omega}=50$.}
	\label{fig15}
\end{figure*}

As the order of resonances increases (\ref{resonance}), the wave interactions become more complex. We begin the analysis of five-wave interactions with the $3\to2$ process:
\begin{equation}
	k_1+k_2+k_3=k_4+k_5,
	\label{3to2}
\end{equation}
since it was first experimentally described in \cite{Ricard21}. The condition (\ref{3to2}) can be written as an equation for  $k_5$:
$$k_5=k_1+k_2+k_3-k_4,$$
which corresponds to the decay of the wave $k_5$ into three co-directional waves $k_1$, $k_2$, and $k_3$ and one wave $k_4$ oppositely directed.
The corresponding fifth-order correlator (quadricoherence) is written as:
\begin{equation}
	Q(k_1,k_2,k_3,k_4)=\frac{|\langle\eta_{k_1}\eta_{k_2}\eta_{k_3}\eta^*_{k_4}\eta^*_{k_1+ k_2+ k_3-k_4}\rangle|}{\sqrt{\langle|\eta_{k_1}\eta_{k_2}\eta_{k_3}|^2\rangle\langle|\eta_{k_4}\eta_{k_1+ k_2+ k_3-k_4}|^2\rangle}} .
	\label{quadr1}
\end{equation}
Figure~\ref{fig13} shows the calculated quadricoherence (\ref{quadr1}) and exact solutions of the system (\ref{resonance}) for the process (\ref{3to2}) for different sets of fixed wavenumbers $k_3$ and $k_4$. One can see that non-trivial five-wave interactions are indeed realized. The non-trivial resonances shown in the figure are found by solving the equation:
$$(k_1+k_2+k_3-k_4)^{3/2}=k_1^{3/2}+k_2^{3/2}+k_3^{3/2}-k_4^{3/2}.$$
For a small value of $k_3$, the trivial intersecting branches of the solution shown in Figure~\ref{fig12}~(a) transform into two distinct lines of non-trivial resonances in Figure~\ref{fig13}~(a). These lines diverge with increase in $k_3$.
It was precisely such resonances that were first reported in \cite{Ricard21}. As $k_4$ approaches $k_3$, the region of resonant interaction narrows, and in the limit $k_4=k_3$, the correlator passes into bicoherence shown in Figure~\ref{fig11}. In the case $k_3>k_4$, the solution (\ref{3to2}) changes significantly, see Figure~\ref{fig14}. Non-trivial resonances are localized near the origin of coordinates, i.e., in the long-wavelength region.

It should be noted that the resonances under consideration (\ref{3to2}) are unlikely to be responsible for the transfer of energy to small scales. The resonances shown in  Figure~\ref{fig13} describe the so-called short-long interactions (the scales of the interacting waves are very different), which are non-local. 
At the same time, the turbulence spectrum (\ref{5w1}) and (\ref{5w2}) is obtained for a direct energy cascade, when the energy passes to small scales as a result of wave disintegration. The energy transition is a local process, i.e., the interactions of waves with close wavenumbers are dominant. The resonances shown in Figure~\ref{fig14}, although local, transfer energy from smaller scales to larger wavelengths.
Thus, the question arises as to which interactions exactly lead to a direct energy cascade.

A possible process that could provide energy transfer to small scales is the disintegration of one wave into four co-directional waves $1\to4$:
\begin{equation}
	k_5=k_1+k_2+k_3+k_4.
	\label{1to4}
\end{equation}
Substitution (\ref{1to4}) into the system of equations  (\ref{resonance}) does not give new non-trivial solutions. Thus, resonances $1\to4$ as well as processes $1\to3$ (\ref{tric2}) are forbidden for plane capillary waves.
For this reason, such resonances are also unlikely to be responsible for the transfer of energy to small scales. 

Finally, we consider another process of wave interaction of  the $3\to2$ type:
\begin{equation}
	k_1+k_2+k_5=k_3+k_4,
	\label{2to3}
\end{equation}
which can be interpreted as the decay of one wave into two pairs of counter-propagating waves: 
\begin{equation}k_5=-k_1-k_2+k_3+k_4.	
\label{2to3n}
\end{equation}
In this representation, the interactions (\ref{3to2}) and (\ref{2to3})  are fundamentally distinct, since they correspond to different sign configurations in the resonance conditions (\ref{resonance}). It is important that the interactions (\ref{2to3n}) have not been considered experimentally \cite{Ricard21}. The quadricoherence corresponding to (\ref{2to3n}) is the following:
\begin{equation}
	Q(k_1,k_2,k_3,k_4)=\frac{|\langle\eta_{k_3}\eta_{k_4}\eta^*_{k_1}\eta^*_{k_2}\eta^*_{-k_1- k_2+ k_3+k_4}\rangle|}{\sqrt{\langle|\eta_{k_3}\eta_{k_4}|^2\rangle\langle|\eta_{k_1}\eta_{k_2}\eta_{-k_1- k_2+ k_3+k_4}|^2\rangle}}.
	\label{quadr3}
\end{equation}
The results of calculations using this formula are presented in Figure~\ref{fig15}. The non-trivial resonance curve is obtained by solving the following algebraic equation:
$$(-k_1-k_2+k_3+k_4)^{3/2}=-k_1^{3/2}-k_2^{3/2}+k_3^{3/2}+k_4^{3/2}.$$
The behavior of the system in Figure~\ref{fig15} differs significantly from that shown in Figures~\ref{fig13}~and~\ref{fig14}. In the present case, the resonant interaction curve is closed and continuously covers a large range of wavenumbers. 
It is important that with an increase in the values of fixed wavenumbers $k_3$ and $k_4$, the region of resonant interaction expands. It is most likely that namely the process (\ref{2to3}) is responsible for the local transfer of energy to small scales, since the processes (\ref{3to2}) either correspond to short-long interactions, Figure~\ref{fig13}, or are limited to large scales, see Figure~\ref{fig14}.

\section{Conclusion}

In this work, a detailed direct numerical simulation of the turbulence of plane-symmetric capillary waves has been carried out. The developed model is fully nonlinear and is based on a dynamic conformal transformation of the region occupied by the fluid into a half-plane of auxiliary conformal variables. The equations of fluid motion are phenomenologically supplemented with terms responsible for the random driving force acting on large scales and the viscosity leading to energy dissipation on short wavelengths.
With this formulation, a direct energy cascade arises, providing a quasi-stationary turbulent state in which the effects of energy pumping are completely compensated by dissipation. The evolution of capillary waves in the wave turbulence regime becomes irregular and chaotic, and the probability density function of the surface amplitude tends to a normal Gaussian distribution. The measured spatial and frequency spectra of turbulence have a power-law form with an exponent close to that obtained under the assumption of the dominant influence of five-wave resonant interactions in anisotropic plane-symmetric geometry, see (\ref{5w1}) and (\ref{5w2}). The measured characteristic nonlinear times are much larger than the linear wave propagation time $\tau_{nl}\gg\tau_l$, which convincingly indicates a weakly nonlinear character of the wave evolution. Performed space-time Fourier analysis did not reveal the presence of any coherent or strongly nonlinear structures. Simulations of freely decaying turbulence allowed us to estimate the value of the Kolmogorov-Zakharov constant. It turned out to be two orders of magnitude greater than the experimentally measured one \cite{Ricard21}, which is closer to the value of the constant for isotropic turbulence. The main result of the work is a high-order correlation analysis of the interaction of plane capillary waves in the weakly turbulent regime. As expected, no non-trivial three- and four-wave resonant interactions are found. Analysis of the results of direct numerical simulation revealed a large number of non-trivial five-wave interactions. We show that the  wave interaction process (\ref{2to3}) is most likely responsible for the local energy transfer to small scales, since the processes  (\ref{3to2}) describe either short-long or large-scale interactions. The wave decay $1\to4$ is forbidden and is not realized in plane-symmetric geometry. In general, the results of direct numerical simulation are in very good agreement with both the weak turbulence theory and the experimental results obtained for plane capillary waves on the mercury surface \cite{Ricard21}.

\begin{acknowledgements}
	The study was supported by a grant from the Russian Science Foundation No. 23-71-10012, https://rscf.ru/en/project/23-71-10012/	
\end{acknowledgements}

\section*{AUTHOR DECLARATIONS}
\section*{Conflict of Interest}
The authors have no conflicts to disclose.
\section*{Author Contributions}
{\bf E. A. Kochurin}: Formal analysis (equal); Funding acquisition (equal);
Investigation (equal); Methodology (equal); Project administration
(equal); Validation (equal); Writing – original draft (equal); Writing –
review \& editing (equal).

{\bf P. A. Russkikh}: Formal analysis (equal); Funding acquisition (equal);
Investigation (equal); Methodology (equal); Project administration
(equal); Validation (equal); Writing – original draft (equal); Writing –
review \& editing (equal).

\section*{DATA AVAILABILITY}
The data that support the findings of this study are available from
the author upon reasonable request.

\end{document}